\documentclass{aa}
\def\sqr#1#2{{\vcenter{\vbox{\hrule height.#2pt
    \hbox{\vrule width.#2pt height#1pt \kern#1pt \vrule width.#2pt}
    \hrule height.#2pt}}}}
\def\dxdy#1#2{{{\partial #1} \over {\partial #2}}}

\usepackage{wasysym}
\usepackage[dvips]{epsfig}
\usepackage{graphicx}
\usepackage{ctable}
\usepackage{natbib}
\usepackage{multirow}
\usepackage[latin1]{inputenc}
\usepackage[T1]{fontenc}
\begin{document}
   \title{Theoretical values of convective turnover times and Rossby numbers 
       for solar-like, pre-main sequence stars%
       \thanks{The complete version of Table\,\ref{tabevol} is only available 
       in electronic form at the CDS via anonymous ftp to cdsarc.u-strasbg.fr 
       (130.79.128.5) or via http://cdsweb.u-strasbg.fr/cgi-bin/qcat?J/A+A/ }}

%   \subtitle{}

   \author{N.R. Landin\inst{1},  
      L.T.S. Mendes\inst{1,2} \and L.P.R. Vaz\inst{1} 
          }

   \offprints{N.R.Landin}

   \institute{Depto.\ de F\'{\i}sica,
              Universidade Federal de Minas Gerais, C.P.702, 31270-901 --
              Belo Horizonte, MG, Brazil; \\
              \email{nlandin@fisica.ufmg.br, lpv@fisica.ufmg.br}
              \and
              Depto.\ de Engenharia Eletr\^onica,
              Universidade Federal de Minas Gerais, C.P.702, 31270-901 --
              Belo Horizonte, MG, Brazil; \\
              \email{luizt@cpdee.ufmg.br}  
             }

   \date{Received \dots; accepted \dots}

% \abstract{}{}{}{}{} 
% 5 {} token are mandatory

\abstract
% context heading (optional)
{
Magnetic fields are at the heart of the observed stellar activity in
late-type stars, and they are presumably generated by a dynamo mechanism
at the interface layer ({\it tachocline\/}) between the radiative core
and the base of the convective envelope. 
}
% aims heading (mandatory)
{
Since dynamo models
are based on the interaction between differential rotation and
convective motions, the introduction of rotation in the \texttt{%
ATON 2.3}
stellar evolutionary code allows for explorations
regarding a physically consistent treatment of magnetic effects in stellar
structure and evolution, even though there are formidable mathematical and
numerical challenges involved. 
}
% methods heading (mandatory)
{
As examples of such explorations, we present theoretical estimates for both
the local convective turnover time ($\tau_{\rm c}$), and global convective
times ($\tau_{\rm g}$) for rotating pre-main sequence
solar-type stars, based on up-to-date input physics for stellar models. Our 
theoretical predictions are compared with the previous ones available in the
literature. In addition, we
investigate the dependence of the convective turnover time on convection
regimes, the presence of rotation and atmospheric 
treatment.  
}
% results heading (mandatory)
{
Those estimates, as opposed to the use of empirically derived values
of $\tau_{\rm c}$ for such matters, can be used to calculate the Rossby
number $Ro$, which is related to the magnetic activity strength in dynamo
theories and, at least for main-sequence stars, shows an observational
correlation with stellar activity. More important, they can also contribute
for testing stellar models against observations.
}
% conclusions heading (optional), leave it empty if necessary 
{Our theoretical values of $\tau_{\rm c}$, $\tau_{\rm g}$ and $Ro$ 
qualitatively agree with those published by Kim \& Demarque (1996). 
By increasing the convection efficiency, $\tau_{\rm g}$ decreases for a given
mass. FST models show still lower values. The presence of rotation 
shifts $\tau_{\rm g}$ towards slightly higher 
values when compared with non-rotating models. The use of non-gray boundary
conditions in the models yields values of $\tau_{\rm g}$ smaller than in the
gray approximation. 
}
%{Place here some concluding remarks?\dots}


\authorrunning {Landin et al.}
\titlerunning {Theoretical Values of Convective Turnover Times and Rossby Numbers}

\maketitle

%________________________________________________________________

\section{Introduction}

Magnetic activity in solar-type stars encompasses a variety of
phenomena, such as star spots, activity cycles, heated outer atmospheres,
X-ray emission, and many others. The driving mechanism for this activity is
generally attributed to a dynamo that results from the interaction between
rotation and convective motions in the star's outer envelope.
Theoretical
work by a number of researchers indicates that for main-sequence, solar-type
stars the field is generated and amplified at the {\it tachocline\/}, the
thin layer of differential rotation between the convection zone and the
nearly rigidly rotating radiative interior.
For stars of a spectral type ranging from mid-F to early-M dwarfs, rotation and
activity are thought to be controlled by this process, also called
an $\alpha$$-$$\Omega$ dynamo \citep{mohanty03}. Its 
efficiency is strongly dependent on the rotation rate and convective timescales.
Young and rapidly rotating stars are, in general, very active. 
Specific models of dynamo theory, such as the $\alpha$$-$$\Omega$ type,
have been successful in explaining the qualitative features
of solar activity \citep{weiss00}. 

Activity is strongly correlated with rotation velocity in the mid-F to mid-M
dwarfs; it increases rapidly with the projected velocity, $v\sin i$, then
saturates above some threshold velocity ($\sim$ 10 km/s). This relationship is
evident only down to K types; as we go from M to M6 types
the rotation-activity connection becomes less clear.
\citet{mohanty03} analyzed rotation velocities and chromospheric
H${\alpha}$ activity, derived from high-resolution spectra, in a sample of 
mid-M to L field dwarfs. They found that, in the spectral type range M4-M8.5,
the saturation-type rotation-activity relation is similar to that in earlier 
types, but the activity saturates at a significantly higher velocity
in the M5.5-M8.5 dwarfs than in the M4-M5 ones; this may result from a change
in the dynamo behavior in later spectral types.

For fully convective stars, such as pre-main sequence late-type stars, this
theory cannot be readily applied, as they miss the tachocline. However, since
magnetic indicators such as active regions and strong flaring have also been
reported for those stars, dynamo mechanisms operating on the full
convection region have also been proposed 
\citep[e.g.][]{durney93}.

On the other hand, observations of stellar
activity in solar-type stars have shown a very tight relationship between
chromospheric Ca II H-K flux and the Rossby number $Ro=P_{\rm rot}/\tau_{\rm c}$,  
where $P_{\rm rot}$ is the rotation period
and $\tau_{\rm c}$ is the local convective turnover time.
The Rossby
number plays an important role in dynamo models, being related to the
{\it dynamo number} $D=Ro^{-2}$ which, in turn, is related to the
growth rate of the field. For solar-type stars, since
$\tau_{\rm c}$ cannot be directly measured, $Ro$ is generally computed through a polynomial
fit of $\tau_{\rm c}$ to the B$-$V color index. For example, \cite{noyes84}
give $\tau_{\rm c}{\rm (B-V)}$ as a theoretically-derived convective overturn time,
calculated assuming a mixing length to scale height ratio $\alpha \sim2$.

For pre-main sequence stars, Rossby numbers have been used to study the
relationship between the X-ray emission and magnetic fields  
\citep[e.g.][]{flacco03c,feigelson03}. In this case,
however, no clear relationship between activity and rotation is seen,
as in the case of main-sequence stars, and one has to resort to evolutionary
models for estimating Rossby numbers.

Though current stellar evolutionary codes are not yet able to deal with
magnetic fields, a first step towards this direction is the introduction
of rotation on the models, as it is a key component of stellar
dynamos. This is the case of the \texttt{ATON 2.3} evolutionary code,
in which both rotation and internal angular momentum redistribution
have been introduced. Such capabilities allow us to make
some exploratory work towards a future version that can handle magnetic
field generation from first principles.

In this work we computed convective turnover times and
Rossby numbers for a range of rotating low-mass stellar models, and
discuss their behavior with time from the 
pre-main sequence to
the zero-age main sequence.
Our results are mainly compared with 
those found by \citet{kim96},
who provided the first self-consistent local 
and global convective turnover times and purely theoretical Rossby numbers
(but see also \citealt{jung2007}).

In Sect. \ref{modinput} we describe the stellar evolutionary 
code used in this work (\texttt{ATON} code) as well as the specific 
inputs adopted for the present calculations. Results on Rossby 
number calculations are discussed in Sect. \ref{calcrossby}. 
The impact of other physical
phenomena (like convection, rotation and atmospheric boundary 
conditions)
is analyzed in Sect. \ref{physimpact}. 
In Sect. \ref{apply} we present a simple application of our
calculations to stars from the IC 2602 open cluster.
Our conclusions are given in Sect. \ref{conclusion}.

\section{Input physics}\label{modinput}

The \texttt{ATON 2.3} code has many updated and modern features regarding the
physics of stellar interiors, of which a full account can be found in
\citet{ventura98}. Some of its most important features are:
most up-to-date OPAL \citep{rogers1} opacities,
supplemented by those of \citet{alexander}
for lower (T $< 6000K$) temperatures; diffusive mixing and overshooting;
and convection treatment under either the mixing length theory (MLT) or the Full
Spectrum of Turbulence (FST) from \citet{canuto91,canuto92} and 
\citet{canuto96}.
In addition to the standard gray atmosphere boundary conditions, the 
\texttt{ATON} code can
also handle non-gray atmospheric integration based on the models
by \citet{heiter,allard1,allard00} as described in
\citet{landin06}.

The structural effects of rotation were implemented
in the \texttt{ATON} code according to the
\citet{kippen70} method, as improved by \citet{endal76}.
Although the \texttt{ATON} code can handle the chemical mixing of species by using a 
diffusive approach (microscopic diffusion), the models presented here use
instantaneous mixing, which is a good approximation for 
pre-main sequence stars \citep{ventura98a}. We account also for the mixing
caused by rotational instabilities \citep{endal78}, including dynamical instabilities
(Solberg-H\o iland and dynamical shear) and secular instabilities (secular 
shear, Goldreich-Schybert-Fricke and meridional circulation).
Angular momentum redistribution in radiative regions was modeled through an
advection-diffusion equation based on the framework of \citet{chaboyer92}
and \citet{zahn92}.
Angular momentum losses in the star's external layers due to magnetized
stellar winds are also taken into account in the form of a boundary
condition at the surface. We adopted
the prescription used in \citet{chaboyer95} with a
``wind index'' $n=1.5$, which reproduces well the \citet{skuma72} law
$v \propto t^{-1/2}$:
{\small
\begin{equation}
    \dxdy{J}{t}\!=\!K \!\left(\!{R\over R_\odot}\!\right)^{\!2-n}\!\!
                  \left(\!{M\over M_\odot}\!\right)^{\!-\frac{n}{3}}\!\!
                  \left(\!{\dot M \over 10^{-14}}\!\right)^{\!1-\frac{2n}{3}}
\!\!\!
                  \omega^{1+\frac{4n}{3}}, \hspace{0.3cm}
                  \omega < \omega_{\rm crit},
\label{eq1}  
\end{equation}
\begin{equation}
    \dxdy{J}{t}\!=\!K\! \left(\!{R\over R_\odot}\!\right)^{\!2-n}
                  \left(\!{M\over M_\odot}\!\right)^{\!-\frac{n}{3}}
                  \left(\!{\dot M \over 10^{-14}}\!\right)^{\!1-\frac{2n}{3}}
\!\!\!
                  \omega \omega_{\rm crit}^{\!\frac{4n}{3}}, \hspace{0.3cm}
                  \omega \geq \omega_{\rm crit},
\label{eq2}
\end{equation}
}
\noindent
where $\omega_{\rm crit}$ introduces a critical rotation level at which
the angular momentum loss saturates (set to $5\,\Omega_\odot$ in
our models, where $\Omega_\odot$ is the current solar surface rotation
rate). The constant $K$ in our models
was calibrated by adjusting a 1\,M$_\odot$ model so that its surface
velocity matches the current solar rotation rate at the equator.
$\dot M$ is the mass loss rate, which enters
in Eqs.\,(\ref{eq1}) and (\ref{eq2}) in units of 
$10^{-14}M_{\odot}\,yr^{-1}$, and which we set to 1.0.

For the initial rotation rates, we adopt the relation between initial angular momentum
(J$_{\rm in}$) and stellar mass obtained
from the corresponding mass-radius and
mass-moment of inertia relations from \citet{kawaler87}:

\begin{equation}
    J_{\rm in} = 1.566 \times 10^{50} \left({M \over M_\odot}\right)^{0.985}
        \quad {\rm g}\,{\rm cm}^2\,{\rm s}^{-1}.
    \label{kaweq}
\end{equation}
For more details about the treatment given to rotation in the
\texttt{ATON} code, see \citet{mendesphd} and \citet{mendes99,mendes03}.

\section{Rossby number calculations}\label{calcrossby}

Convective turnover times and Rossby numbers were computed for models
ranging from 0.6 to 1.2\,M$_\odot$ (in 0.1\,M$_\odot$ increments) 
with solar chemical composition.
For ease of comparison with a previous work by 
\citet{kim96}, who provided theoretical calculations of 
Rossby numbers for pre-main sequence stars,
convection was treated according to the MLT, with the free parameter
$\alpha$ set to 1.5 (which fits the solar radius at the solar age for a 
gray atmosphere, non-rotating model). Convection treatment and boundary conditions
considerably affect the calculations of global convective turnover times
and, consequently, the Rossby numbers (see Sect. \ref{convsubsec}).
Rotation was modeled according to rigid body law in
convective zones and local conservation of angular momentum in radiative regions
\citep{mendes03}.

We followed the evolution of local and global convective turnover times (to be defined shortly) and
Rossby numbers during the pre-main sequence for 0.6-1.2\,M$_{\odot}$ stars
and tabulated them together with the corresponding evolutionary tracks.
Table\,\ref{tabevol} presents the 1\,M$_{\odot}$ models as an example of
such tables. Column 1 gives the logarithm of stellar age (in years); 
Col. 2 the logarithm of stellar luminosity (in solar units);
Col. 3 the logarithm of effective temperature (in K); 
Col. 4 the logarithm of effective gravity (in cgs);
Col. 5 the logarithm of local convective turnover time (in seconds);
Col. 6 the global convective turnover time (in days);
and Col. 7 the Rossby number.

\begin{figure}[tb]
   \centering{
   \includegraphics[width=9cm]{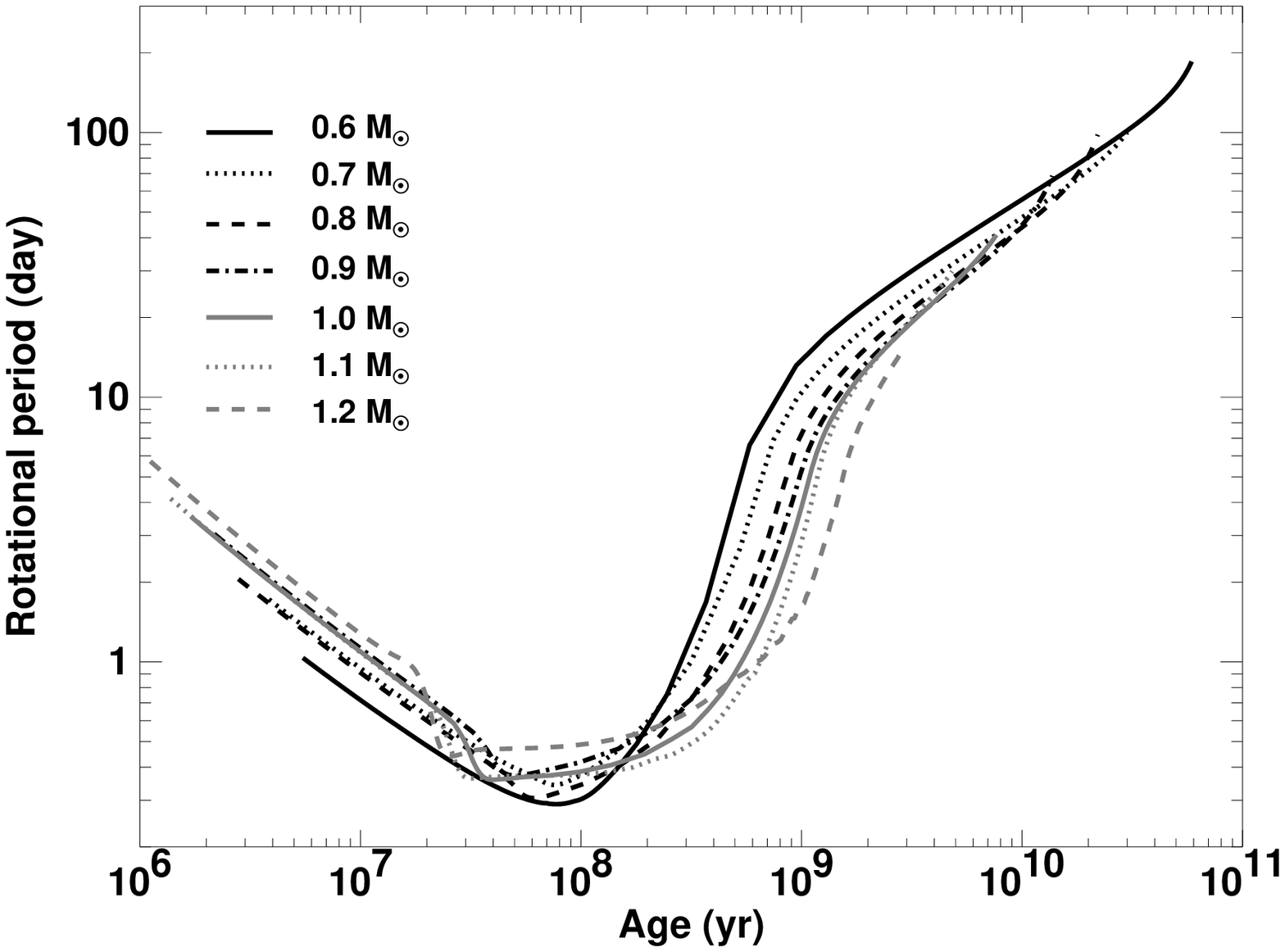}
             }
   \caption[P$_{\rm rot}$ vs. age.]
            {\textmd{Rotation period plotted against age, for each 
            model mass.
            }}
   \label{prxageleft}
\end{figure}

\begin{table}[h]
\caption{Pre-main sequence evolutionary tracks (including $\log \tau_c$,
$\log\tau_g$ and $Ro$) for 1\,M$_{\odot}$ star$^a$.
}
%-------------------------------------
% Table 1
%-------------------------------------
%
\vspace{0.2cm}
\label{tabevol}
\centering
\advance\tabcolsep by -4pt
\begin{tabular}{rrrrrrc}
\hline \hline
${\log Age\atop{\rm (years)}}$ & $\log\frac{L}{L_{\odot}}$ & ${\log T_{\rm eff}\atop{\rm (K)}}$ & $\log g\atop{\rm (cgs)}$ & $\log \tau_c\atop{\rm (seconds)}$ & $\tau_g\atop\rm {(days)}$ & $Ro$ \\ \hline
  6.2324 &    0.1468 & 3.6524 & 3.855 & 6.3275 & 766.6288 & 0.1436 \\ [-1.5pt]
  6.4053 &    0.0217 & 3.6513 & 3.976 & 7.1018 & 488.2106 & 0.0183 \\ [-1.5pt]
  6.5840 & $-$0.1029 & 3.6499 & 3.095 & 7.0810 & 369.1187 & 0.0146 \\ [-1.5pt]
  6.7666 & $-$0.2172 & 3.6494 & 4.207 & 7.0074 & 291.4590 & 0.0131 \\ [-1.5pt]
  6.9324 & $-$0.2965 & 3.6522 & 4.297 & 6.9087 & 227.1363 & 0.0129 \\ [-1.5pt]
  7.0786 & $-$0.3278 & 3.6606 & 4.362 & 6.7949 & 179.1020 & 0.0135 \\ [-1.5pt]
  7.2016 & $-$0.2987 & 3.6775 & 4.401 & 6.6878 & 134.8216 & 0.0145 \\ [-1.5pt]
  7.2978 & $-$0.2231 & 3.6999 & 4.415 & 6.5727 &  99.6126 & 0.0165 \\ [-1.5pt]
  7.3740 & $-$0.1262 & 3.7232 & 4.411 & 6.4301 &  72.7130 & 0.0204 \\ [-1.5pt]
  7.4394 & $-$0.0387 & 3.7445 & 4.409 & 6.2546 &  53.4192 & 0.0269 \\ [-1.5pt]
  7.5214 & $-$0.0924 & 3.7525 & 4.494 & 6.1771 &  41.6650 & 0.0232 \\ [-1.5pt]
  7.9918 & $-$0.1590 & 3.7461 & 4.536 & 6.1942 &  46.0585 & 0.0212 \\ [-1.5pt]
  9.0247 & $-$0.1138 & 3.7531 & 4.518 & 6.1821 &  42.4793 & 0.2593 \\ [-1.5pt]
  9.3023 & $-$0.0862 & 3.7553 & 4.499 & 6.1523 &  41.0925 & 0.7916 \\ [-1.5pt]
  9.4669 & $-$0.0580 & 3.7572 & 4.479 & 6.1598 &  39.4493 & 1.0789 \\ [-1.5pt]
  9.5829 & $-$0.0288 & 3.7591 & 4.457 & 6.1303 &  38.1493 & 1.4206 \\ [-1.5pt]
  9.6573 & $-$0.0040 & 3.7605 & 4.438 & 6.1369 &  38.0850 & 1.5989 \\ [-1.5pt]
  9.6980 &    0.0119 & 3.7612 & 4.425 & 6.1409 &  37.5359 & 1.7020 \\ [-1.5pt]
  9.7644 &    0.0428 & 3.7624 & 4.399 & 6.1102 &  36.8616 & 2.0771 \\ [-1.5pt]
  9.8211 &    0.0754 & 3.7634 & 4.370 & 6.1166 &  36.2118 & 2.3108 \\ \hline
\multicolumn{7}{p{0.95\columnwidth}}{$^a$The complete version of the table, including
seven tracks for all the masses of Table\,\ref{table-isoch2}, will be
available in electronic form.}
\end{tabular}
\end{table}
Figure~\ref{prxageleft} depicts the rotation period as a function of age for
all models, and shows the typical spin-up during contraction followed by
the longer phase of continuous spin-down.
For the 1\,M$_{\odot}$ model, this results in an initial velocity of nearly
3\,km\,s$^{-1}$ at the beginning of the Hayashy phase; this is about one order
of magnitude below the value used by \citet{kim96}.
When calculating convective overturn times, one must make an arbitrary
assumption about where in the convection zone the dynamo is operating.
This assumption significantly affects the value of $\tau_{\rm c}$,
since it is strongly dependent on the depth. In this work, we follow \cite{gilliland}
and calculate the ``local'' convective turnover time $\tau_{\rm c}$ at a distance of one-half the mixing
length, $\alpha\,H_{\rm P}/2$, above the base of the convection zone.
Its value is computed through the equation 
$\tau_{\rm c} = \alpha\,H_{\rm P}/v$, where $v$ is the convective velocity.
It should be noted, however, that there is no agreement in the literature
regarding the precise location where $\tau_{\rm c}$ should be computed; \cite{montesinos01}, for example,
calculate it at distance of $0.95\,H_{\rm P}$ above the base of convective zone, with $H_{\rm P}$ taken
as its value at the base of the convection zone. 

\begin{figure}[tb]
   \centering{
   \includegraphics[width=9cm]{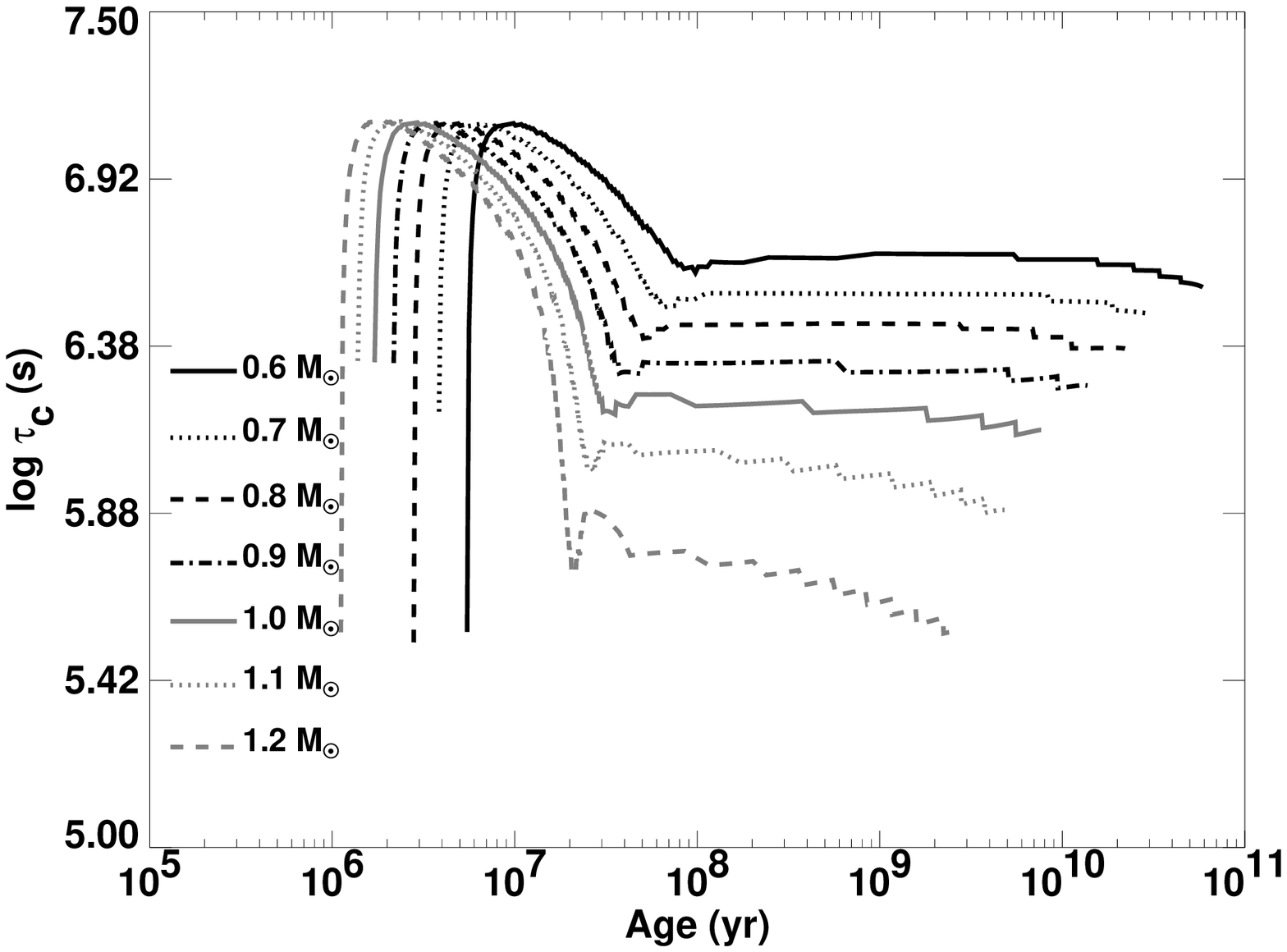} 
             }
   \caption[Local $\tau_{\rm c}$ vs. age.]
           {\textmd{``Local'' convective 
            turnover time as a function of age for all models.
	    }}
   \label{prxageright}
\end{figure}

\begin{figure}[tb]
   \centering{
  \includegraphics[width=9cm]{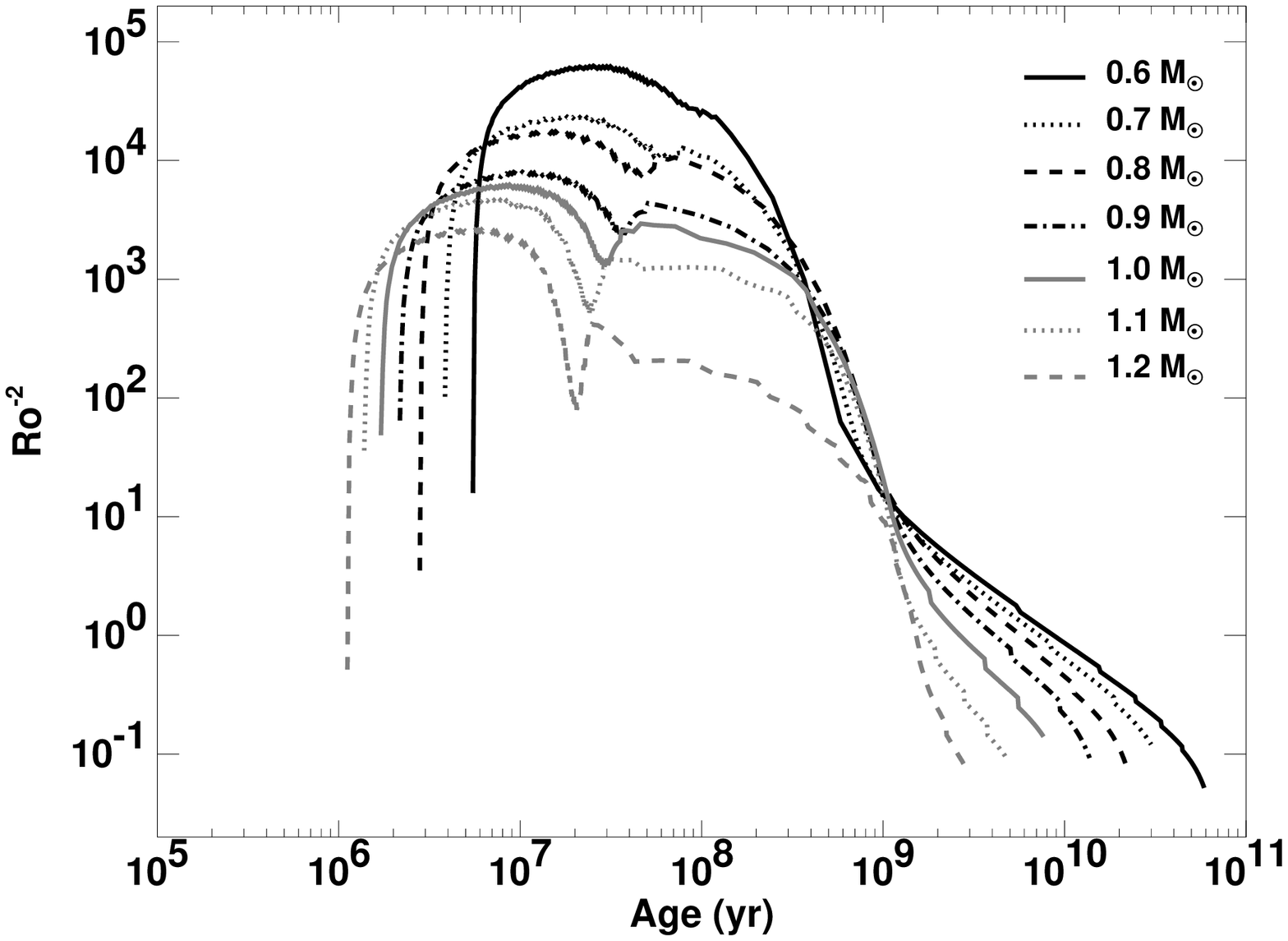} 
              }
   \caption[Dynamo number vs. age.]
            {\textmd{The dynamo number
            as a function of age for each model mass.
	    }}
   \label{taugcxageright}
\end{figure}

Figure~\ref{prxageright} shows
the time evolution of the local convective turnover time 
(in seconds).
For a given mass, it decreases during the Hayashy contraction and reaches its minimum when the
contraction stops. After that, $\tau_{\rm c}$ remains constant until the stars reach
a main sequence configuration.
\begin{figure}[tb]
   \centering{
   \includegraphics[width=9cm]{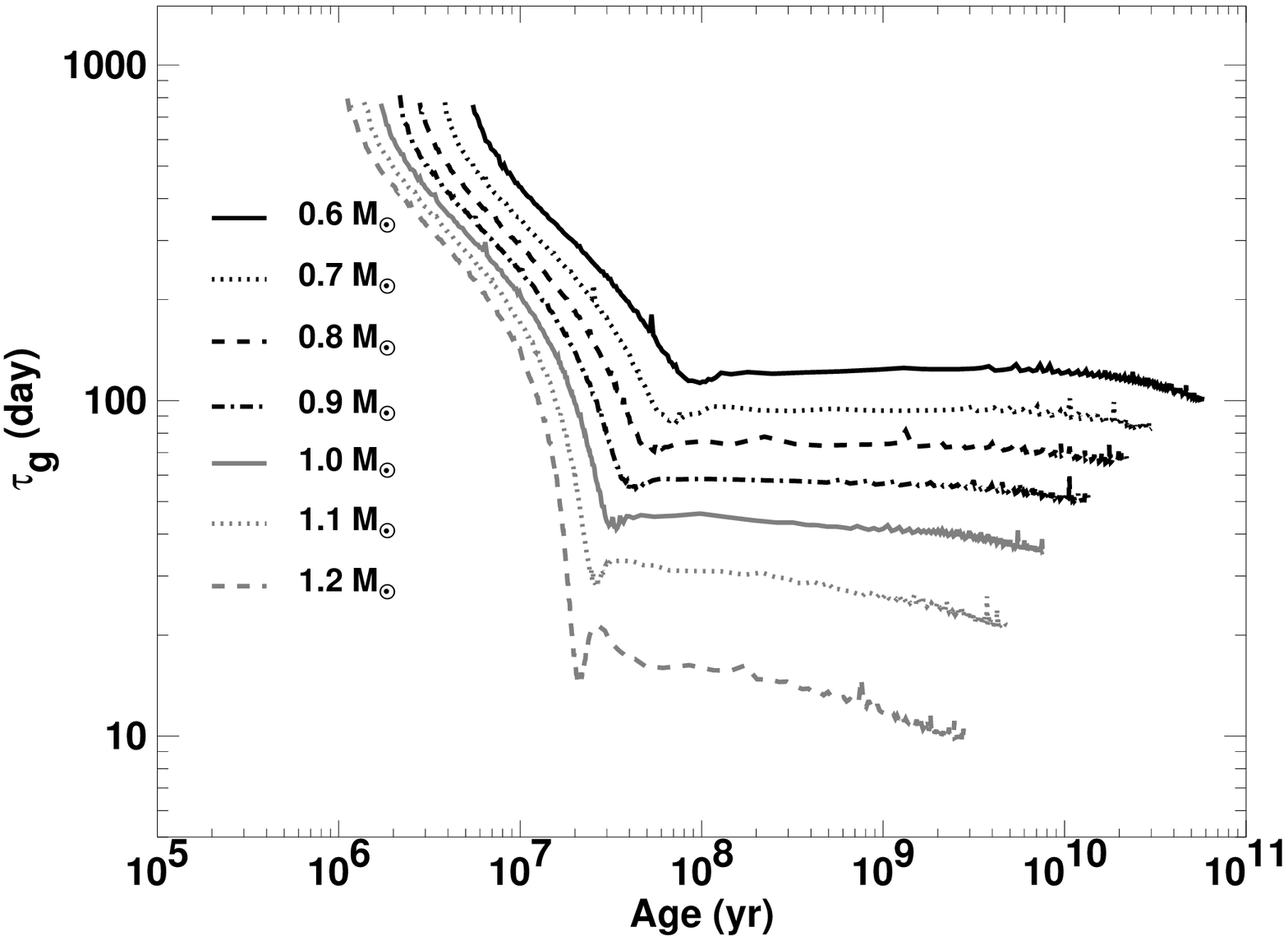} 
              }
   \caption[Global convective turnover time vs. age.]
            {\textmd{``Global''  
            convective turnover time 
            as a function of age for each model mass.
	    }}
   \label{taugcxageleft}
\end{figure}
However, most relevant for our purposes are 
Figs.~\ref{taugcxageleft} and \ref{taugcxageright}, which show the profiles of the ``global'' convective
turnover time and the ``dynamo number'', $Ro^{-2}$, respectively, 
as functions of age. The global convective turnover time ($\tau_{\rm g}$) 
is defined as 

\begin{equation}
    \tau_{\rm g} = \int_{R_b}^{R_\star} {dr \over v},
\end{equation}

\noindent 
where $R_b$ is the radius at the bottom of the convective zone, $R_\star$
is the stellar radius and $v$ is convective velocity.

Figure~\ref{taugcxageleft} shows that $\tau_{\rm g}$ follows the same behavior
as the local convective turnover time, also decreasing 
substantially during contraction to the zero-age main sequence and, after 
that, remaining nearly constant and depending only on the mass. 
As in \cite{kim96}, the local convective turnover time behaves like $\tau_{\rm g}$ except for
a scaling factor, because the convective turnover timescale is weighted towards the deepest
part of the convection zone, where the shortcomings of the mixing length approximation
are less important.
With regard to $Ro^{-2}$, it is seen from Fig.~\ref{taugcxageright}
that it follows $\tau_{\rm c}$ during contraction but, after that, 
decreases as expected since the rotation period also increases.

By using the evolutionary tracks, we constructed a set of isochrones
for the ages of 0.2, 0.5, 0.7, 1.0, 2.0, 4.55 (solar age), 10 and 15 Gyr. 
In Table~\ref{table-isoch2} we list their characteristics such as
stellar mass in solar masses (Col. 1); 
logarithm of the effective temperature, in Kelvin 
(Col. 2); 
logarithm of the stellar luminosity, in cgs (Col. 3); 
global convective turnover time, in days 
(Col. 4); dynamo number (Col. 5); and the rotation period, in days (Col. 6). 

In Fig.~\ref{isotauteff} we show a plot of the global and local convective turnover times
versus $\log(T_{\rm eff})$ for each {age. 
Isochrones for the global convective turnover time versus period are shown in 
Fig.~\ref{isotauper}. The highest point of each curve 
corresponds to the lowest mass. For a given mass, $\tau_{\rm g}$ varies slightly with the period.
For a given period and age, the global convective turnover time 
depends on stellar mass.

\begin{table}[tb]
\caption{Isochrones for all models.} 
\label{table-isoch2}
\centering
{\footnotesize
\advance\tabcolsep by -3pt
\begin{tabular}{rrrrrr}
\hline \hline
Mass         &                  &                      & $\tau _{\rm g}$~~~~&             & Rotation \\ 
(M$_{\odot}$)& $\log$ (T$_{\rm eff}$) & $\log$ (L/L$_{\odot}$) &  (days )  & $Ro^{-2}$ & Period (d)\\ \hline
\multicolumn{6}{c}{0.2\,Gyr}                                          \\ [-1.5pt]
\hline
0.60 & 3.6148 & $-$1.13663 & 121.1802 & 8627.8177 & 0.5581 \\ [-1.5pt]
0.70 & 3.6508 & $-$0.85884 &  93.8181 & 4711.2816 & 0.5913 \\ [-1.5pt]
0.80 & 3.6878 & $-$0.60171 &  77.0336 & 4601.9900 & 0.4777 \\ [-1.5pt]
0.90 & 3.7218 & $-$0.36391 &  57.9367 & 2093.6635 & 0.5387 \\ [-1.5pt]
1.00 & 3.7480 & $-$0.14791 &  43.8863 & 1643.1720 & 0.4534 \\ [-1.5pt]
1.10 & 3.7695 &    0.05200 &  30.5734 &  875.9235 & 0.4187 \\ [-1.5pt]
1.20 & 3.7908 &    0.23895 &  14.8697 &  136.2882 & 0.5509 \\ [-1.5pt]
\hline
\multicolumn{6}{c}{0.5\,Gyr}                                          \\ [-1.5pt]
\hline
0.60 & 3.6148 & $-$1.12952 & 121.9241 &  365.2809 &  3.3443 \\ [-1.5pt]
0.70 & 3.6511 & $-$0.85206 &  94.4531 &  364.8090 &  2.2956 \\ [-1.5pt]
0.80 & 3.6898 & $-$0.59006 &  73.5144 &  569.3372 &  1.3836 \\ [-1.5pt]
0.90 & 3.7241 & $-$0.35056 &  57.1289 &  426.6247 &  1.2034 \\ [-1.5pt]
1.00 & 3.7511 & $-$0.13146 &  42.1241 &  365.2252 &  0.9081 \\ [-1.5pt]
1.10 & 3.7734 &    0.07207 &  28.4896 &  268.6685 &  0.7299 \\ [-1.5pt]
1.20 & 3.7938 &    0.26043 &  13.4697 &   42.9852 &  0.8630 \\ [-1.5pt]
\hline
\multicolumn{6}{c}{0.7\,Gyr}                                          \\ [-1.5pt]
\hline
0.60 & 3.6146 & $-$1.12785 & 122.7990 &   45.4688 &  8.1315 \\ [-1.5pt]
0.70 & 3.6512 & $-$0.84948 &  93.9007 &   69.1059 &  5.3534 \\ [-1.5pt]
0.80 & 3.6903 & $-$0.58622 &  73.8996 &  130.3651 &  2.8620 \\ [-1.5pt]
0.90 & 3.7248 & $-$0.34541 &  56.8609 &  115.0898 &  2.1514 \\ [-1.5pt]
1.00 & 3.7521 & $-$0.12448 &  42.0428 &  122.1356 &  1.5846 \\ [-1.5pt]
1.10 & 3.7748 &    0.08167 &  26.7088 &   92.5457 &  1.1571 \\ [-1.5pt]
1.20 & 3.7950 &    0.27275 &  12.7590 &   23.7916 &  1.0832 \\ [-1.5pt]
\hline
\multicolumn{6}{c}{1\,Gyr}                                          \\ [-1.5pt]
\hline
0.60 & 3.6145 & $-$1.12602 & 123.9869 &   15.6947 & 13.7087 \\ [-1.5pt]
0.70 & 3.6514 & $-$0.84667 &  93.3632 &   15.1359 & 10.3630 \\ [-1.5pt]
0.80 & 3.6908 & $-$0.58179 &  74.2237 &   20.3958 &  7.2070 \\ [-1.5pt]
0.90 & 3.7255 & $-$0.33908 &  56.5470 &   19.0868 &  5.2591 \\ [-1.5pt]
1.00 & 3.7530 & $-$0.11547 &  41.4909 &   21.0411 &  3.8355 \\ [-1.5pt]
1.10 & 3.7760 &    0.09436 &  25.8848 &   15.9412 &  2.8429 \\ [-1.5pt]
1.20 & 3.7962 &    0.28993 &  11.7507 &    9.2329 &  1.6338 \\ [-1.5pt]
\hline
\multicolumn{6}{c}{2\,Gyr}                                          \\ [-1.5pt]
\hline
0.60 & 3.6149 & $-$1.11995 & 124.0207 &    5.5953 & 22.8118 \\ [-1.5pt]
0.70 & 3.6523 & $-$0.83796 &  93.6465 &    4.6988 & 18.5225 \\ [-1.5pt]
0.80 & 3.6925 & $-$0.56838 &  74.4090 &    4.3392 & 15.5263 \\ [-1.5pt]
0.90 & 3.7275 & $-$0.31932 &  56.1457 &    2.8531 & 13.6315 \\ [-1.5pt]
1.00 & 3.7553 & $-$0.08640 &  41.0063 &    1.6047 & 12.9745 \\ [-1.5pt]
1.10 & 3.7783 &    0.13614 &  24.9018 &    0.6022 & 12.7151 \\ [-1.5pt]
1.20 & 3.7979 &    0.34593 &  10.3207 &    0.2400 &  9.0570 \\ [-1.5pt]
\hline
\multicolumn{6}{c}{4.55\,Gyr}                                       \\ [-1.5pt]
\hline
0.60 & 3.6163 & $-$1.10689 & 123.5031 &    2.1587 & 36.6458 \\ [-1.5pt]
0.70 & 3.6548 & $-$0.81720 &  93.2665 &    1.7031 & 30.6947 \\ [-1.5pt]
0.80 & 3.6970 & $-$0.53436 &  71.9753 &    1.3278 & 27.0554 \\ [-1.5pt]
0.90 & 3.7327 & $-$0.26633 &  56.5844 &    0.8790 & 24.8868 \\ [-1.5pt]
1.00 & 3.7605 & $-$0.00376 &  38.2314 &    0.3903 & 25.3930 \\ [-1.5pt]
1.10 & 3.7807 &  0.25629 &  21.7054 &    0.1036 & 27.8899 \\ [-1.5pt]
\hline
\multicolumn{6}{c}{10\,Gyr}                                          \\ [-1.5pt]
\hline
0.60 & 3.6196 & -1.08068 & 121.1537  &   0.8633 & 55.7920 \\ [-1.5pt]
0.70 & 3.6610 & -0.77093 &  90.9166  &   0.6376 & 47.6948 \\ [-1.5pt]
0.80 & 3.7070 & -0.45253 &  69.9406  &   0.4504 & 43.8041 \\ [-1.5pt]
0.90 & 3.7428 & -0.12491 &  51.0290  &   0.2126 & 44.6414 \\ [-1.5pt]
\hline
\multicolumn{6}{c}{15\,Gyr}                                          \\ [-1.5pt]
\hline
0.60 & 3.6228 & $-$1.05579 & 119.3184  &   0.5619 & 69.1454 \\ [-1.5pt]
0.70 & 3.6674 & $-$0.72295 &  89.8339  &   0.3972 & 60.0909 \\ [-1.5pt]
0.80 & 3.7163 & $-$0.35931 &  66.2647  &   0.2143 & 58.7162 \\ [-1.5pt] \hline
\end{tabular}
}
\end{table}

\begin{figure}[tb]
   \centering{
   \includegraphics[width=9cm]{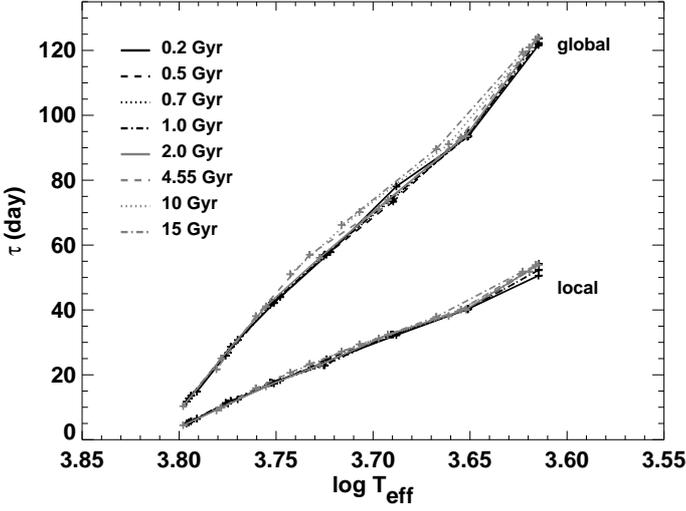}
             }
   \caption{The global and local convective turnover times as a function of 
effective temperature and age. 
            }
   \label{isotauteff}
\end{figure}

\begin{figure}[htb]
   \centering{
   \includegraphics[width=9cm]{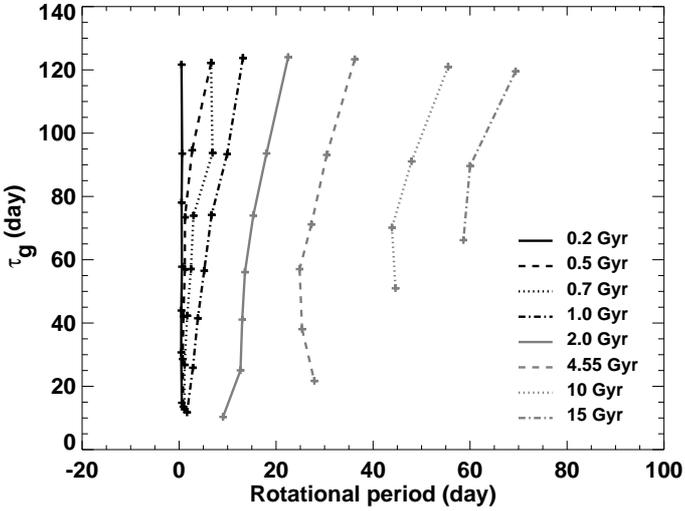} 
             }
   \caption{The global convective turnover time as a function of 
            rotation period and age. 
            }
   \label{isotauper}
\end{figure}

Figure~\ref{isoper} shows the rotation period
versus $\log(T_{\rm eff})$. In that figure the rightmost points 
(those with the lowest temperatures) correspond to the lowest mass
considered, namely 0.6\,M$_{\odot}$. We recall that in our rotating models convective regions
are modeled by assuming rigid body rotation, while radiative regions
rotate according to local conservation of angular momentum.

In Fig.~\ref{isodyn}
we plotted the inverse square of the Rossby number
versus the rotation period. The lowest point of each line
represents the highest mass. 
\begin{figure}[tb]
   \centering{
   \includegraphics[width=9cm]{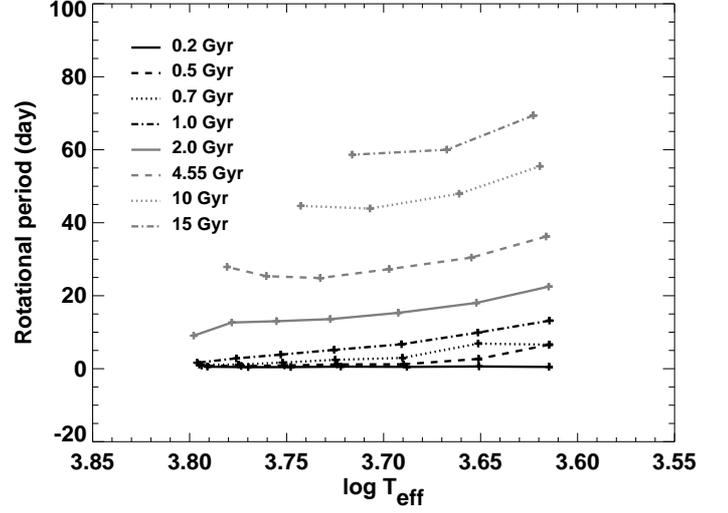}
             }
   \caption{Rotation period as a function of effective temperature and age.
            }
   \label{isoper}
\end{figure}
\begin{figure}[tb]
   \centering{
   \includegraphics[width=9cm]{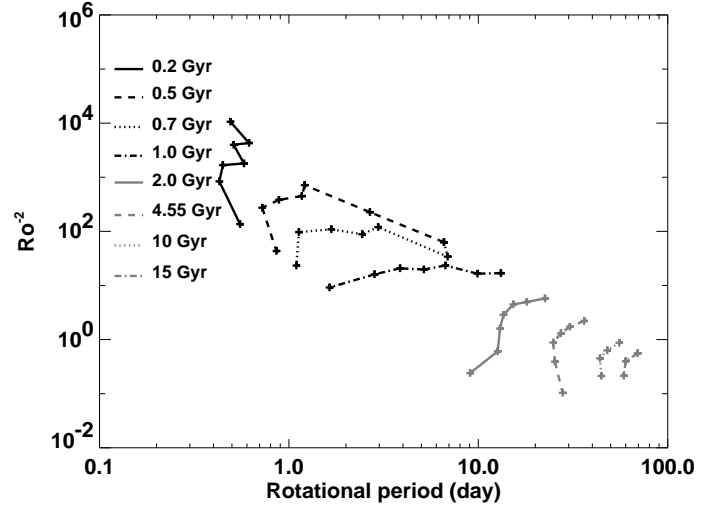} 
             }
   \caption{The dynamo number as a function of rotation period and age.
            }
   \label{isodyn}
\end{figure}
Figure~\ref{isondytef} shows the dynamo number
$Ro^{-2}$ as a function of effective temperature and age. 
After establishing an empirical relation between $Ro^{-2}$ and magnetic activity indices,
one can use Fig.~\ref{isondytef} to estimate the stellar mass and age from the effective
temperature and an activity index.

We believe that this set of results can be useful to support observational
studies of active pre-main sequence stars as well as for testing
stellar models against observations.
Of course, these determinations are model-dependent and subjected to 
the assumptions made in each model. 

When compared to the work of \cite{kim96}, our results concerning
both the global and local convective turnover times agree well with
theirs; but our Rossby number values are in general lower by one order of magnitude.
Since $Ro=P_{\rm rot}/\tau_{\rm c}$,  this difference can be mainly attributed to the
different initial angular momentum of the models: while we use values of $J_{\rm in}$
from Eq. \ref{kaweq}, which are different for each stellar mass, \citet{kim96}
adopted an initial rotation velocity of 30 km s$^{-1}$ for all their models, which
results in higher initial rotation rates (for example, the initial rotation
velocity for a 1 M$_\odot$ model corresponds to near 3 km s$^{-1}$ in our 
case). 
Some other factors can also contribute to the observed 
differences in $Ro$; for example, although
convection is treated according to MLT approximation in both models,
the convection efficiency is slightly different: while
we adopt $\alpha_{\rm MLT}$=1.5, \citet{kim96}
use $\alpha_{\rm MLT}$=1.86315; these values correspond to the $\alpha$ 
which reproduce the solar radius at the solar age in each model.
In addition, the opacities used in our models \citep{rogers1,alexander} are more up-to-date
than those used by \citet{kim96}, namely those by \citet{rogers91} and
\citet{kurucz91}.
However, their atmospheric boundary conditions \citep{kurucz92} are 
more realistic than the gray atmosphere approximation used in our work. 
Finally, the details
concerning the internal redistribution of angular momentum can also 
be a source of such differences.

Stellar activity is a complex phenomena rooted on the intricate
interactions between magnetic fields, convection and differential rotation.
In the case of the Sun and low-mass stars, it 
is generally accepted that magnetic fields are produced by a dynamo process inside
the stars.

Though stellar models that include rotation, such as the
\texttt{ATON 2.3} and the one from \citet{kim96}, can indeed compute
purely theoretical Rossby numbers, it is clear that a more consistent scenario
could be obtained if magnetic field generation was included in the models.
For example, since the Rossby number $Ro$ is closely related to the so-called
``dynamo number'' which plays a central role in the $\alpha\Omega$-dynamos,
stellar models that include a dynamo process would allow a cross-check between
these two values. This is however a difficult task, though some initiatives have
been reported in the literature (e.g. \citealt{lydon95}, \citealt{li06}). As a
matter of fact, work is in progress  to include the effects of magnetic fields in
the \texttt{ATON 2.3} code itself, which we plan to address
in a forthcoming paper.

Asteroseismological data, which now are increasingly available through
missions such as CoRoT and Kepler, are obviously of great importance to impose additional
constraints on rotating stellar models,
such as convective zone depth, rotation rate and differential rotation
(see e.g. \citealt{dupret04}, \citealt{christensen08}, and
\citealt{tang08}).

\begin{figure}[tb]
   \centering{
   \includegraphics[width=9cm]{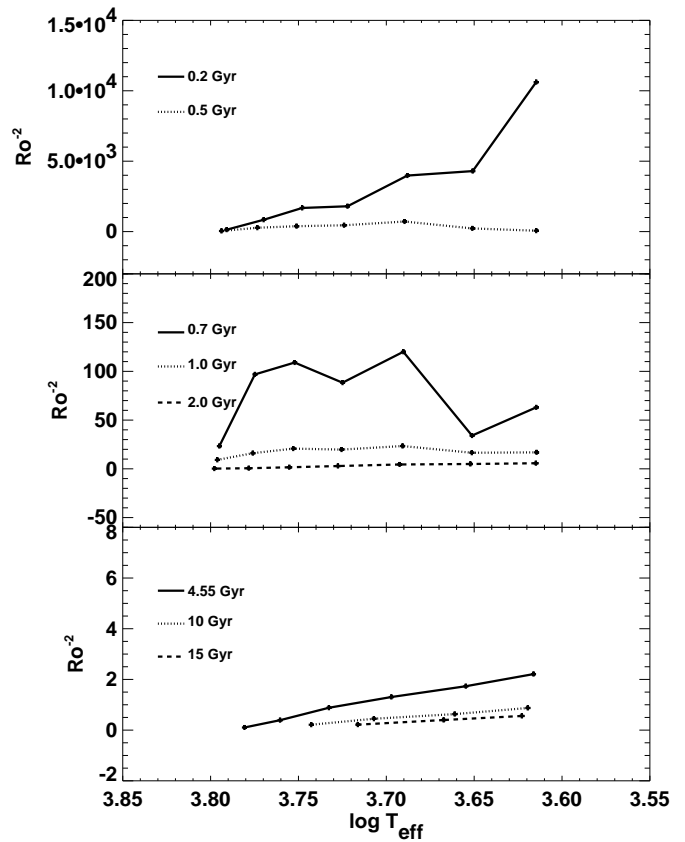} 
             }  
   \caption[$Ro^{-2}$ vs. T$_{\rm eff}$ for some isochrones.]{Dynamo number 
versus $\log(T_{\rm eff})$ and age for isochrones of 0.2, 0.5, 0.7, 
            1.0, 2.0, 4.55 (solar age), 10 and 15 Gyr.}
   \label{isondytef}
\end{figure}

\section{The impact of some physical phenomena on convective turnover times}
\label{physimpact}

\subsection{Convection}\label{convsubsec}

In principle, a correct calculation of convective turnover times depends on
a good knowledge about stellar convection. Unfortunately, our poor 
understanding of this subject limits all determinations 
of such time scales.

Presently, there are three main ways of computing convection in stellar 
envelopes: i) the traditional Mixing Length Theory \citep[MLT,][]{bohm58};
ii) the Full Spectrum of Turbulence \citep[FST,][]{canuto96}; and iii) MLT, in which
the $\alpha$ value for each gravity and $T_{\rm eff}$ is calibrated
upon 2D or 3D hydrodynamical simulations \citep{ludwig99,ludwig2002}.

Since convective turnover times are strongly affected by 
the convection treatment, we analyze their dependence with 
different convection models and different convection efficiencies. 
With our version of \texttt{ATON}, we can
compute convective turnover times by using two different convection
regimes, the classical MLT and the FST. 
The other physical parameters which do not concern convection,
remain as described in Sect.~\ref{modinput}.

In Fig.\,\ref{taugconv} we show $\log \tau_{\rm g}$
as a function of stellar age for 0.5, 0.7, 1.0 and 1.2\,M$_{\odot}$
models computed with FST or MLT, with $\alpha$=1.0, 1.5 and 2.0 for the latter case. 
After the radiative core is formed, the global convective turnover time 
decreases until the pre-main sequence contraction is over;
the decrease of $\tau_{\rm g}$ gets sharper as the stellar mass
increases, reflecting the receding size of the convective region.
During the first stages of main sequence evolution $\tau_{\rm g}$ 
remains nearly constant.
For MLT models, the higher the convection efficiency 
(larger $\alpha$ parameter), the lower $\tau_{\rm g}$ and hence a higher 
Rossby number; 
however, for masses above 1\,M$_{\odot}$, the situation is 
reversed as the star enters the main sequence. The FST models show 
still lower values of $\tau_{\rm g}$, as it describes the whole spectrum 
of convective eddies and consequently convective velocities are computed 
in an intrinsically different way.

The results obtained for the rotating, pre-main sequence models
show the same behavior obtained elsewhere for non-rotating, main
sequence stars \citep[e.g.][]{pizzolato01} regarding the chosen 
convection model. 

\begin{figure}[tb]
\centering{
\includegraphics[width=9cm]{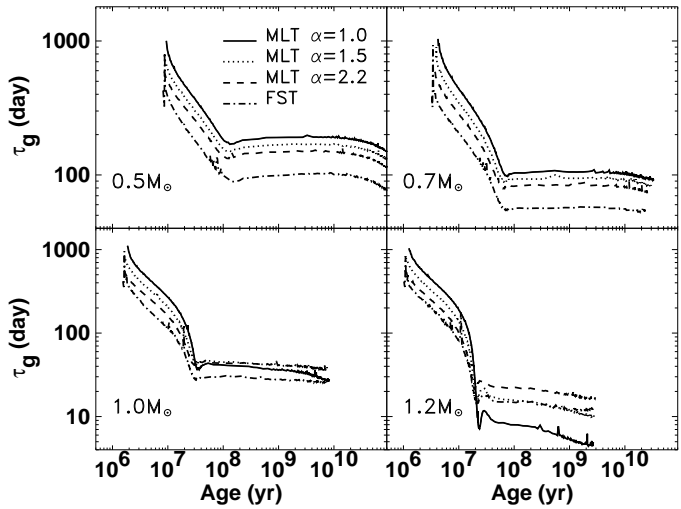}
}
\caption{Global convective turnover time as a function of age 
for selected models, for different masses and convective regimes.
}
\label{taugconv}
\end{figure}

\subsection{Rotation}\label{rotation}

According to \citet{durney79}, rotation has a stabilizing effect on 
convection since it reduces the growth rate of instability. Even in the Sun,
which has a low surface angular velocity, rotation can indeed influence 
convection.

In Fig.~\ref{taugrot} we show the global convective turnover time for
0.5, 0.7, 1.0 and 1.2 M$_{\odot}$, calculated with non-rotating models (solid
lines),
with solid body rotation throughout the whole star (SB, dotted lines) and 
with a combination of
solid body rotation in convective regions and differential rotation in
radiative zones (SB+Diff, dashed lines).
Parameters related to other physical inputs remain as in 
Sect.~\mbox{\ref{modinput}}}.
In the case of SB rotation, the whole star rotates with the
same angular velocity during the 
evolution, while in the SB+Diff case our models take into account the
surface angular momentum loss from stellar winds and the internal redistribution
of angular momentum.
Since our models correspond to slowly rotating stars (for example, our 1 M$_{\odot}$
models have an initial velocity of $\sim$3 km/s
at the beginning of the Hayashy 
phase), the influence of rotation on the global convective turnover time becomes smaller for lower masses,
being more evident at the end of the Hayashy phase and for stars with
masses larger than 1 M$_{\odot}$. 
The values of $\tau_{\rm g}$ obtained 
with models considering SB rotation are higher than those resulting from
non-rotating models and models with SB+Diff rotation.
Higher values of $\tau_{\rm g}$ in the presence of rotation are indeed
expected, since one of the main effects of rotation is to mimic a lower mass star with
a larger convective envelope. In addition, when comparing values of $\tau_{\rm g}$ for
SB rotation and for SB+Diff rotation, one would also expect lower values for the latter since
in this case the angular velocity at the convection zone is lower due to the loss of angular momentum
by winds and the redistribution of angular momentum.

During the main-sequence evolution, the values of $\tau_{\rm g}$ for SB+Diff rotation
approaches the non-rotating ones as the rotation rate decreases, as can be seen
in Fig.~\ref{taugrot} (dashed and solid curves). 
It is worth noting
that these results consider the same initial angular momentum for SB and SB+Diff rotation, as given by Eq. ~(\ref{kaweq}).
Some tests made with 1~M$_{\odot}$ models reveal that $\tau_{\rm g}$
increases with the initial angular velocity (and so the initial 
angular momentum).

\begin{figure}[tb]
\centering{
\includegraphics[width=9cm]{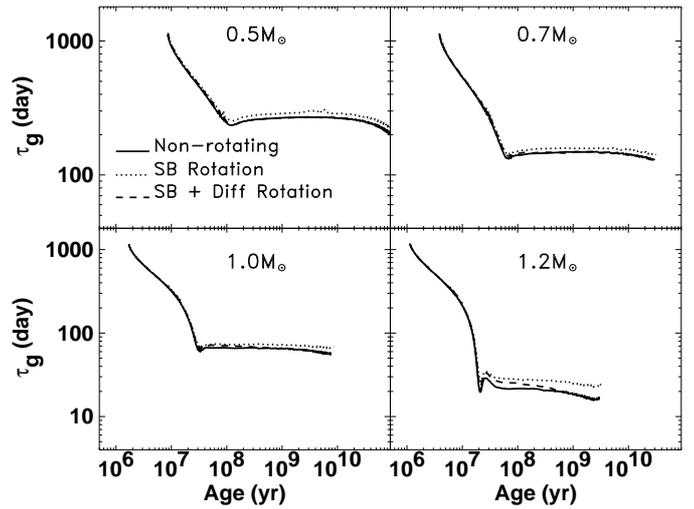}
}
\caption{Global convective turnover time as a function of age 
for selected models, for different masses and rotation regimes.
}
\label{taugrot}
\end{figure}

\subsection{Non-gray atmosphere}\label{ngatm}

There is a good agreement in the literature that the use of non-gray atmosphere models
as boundary conditions for stellar structure and evolution codes produces
evolutionary tracks shifted toward lower effective temperatures in 
HR diagram. 
The non-gray models include the treatment of atmospheric convection,
which cannot be neglected at low T$_{\rm eff}$'s. The use of 
frequency-dependent opacities may also modify the onset
of convection within the atmosphere, so that the T$_{\rm eff}$ of the evolutionary 
tracks can be strongly affected. Non-gray models produce stars with
larger and deeper convective envelopes when compared with gray models, though
their stellar radius is smaller.

Here we report values of $\tau_{\rm g}$ calculated with 
non-rotating models
which use either gray or non-gray atmosphere 
boundary conditions. 
The other input physics are as in Sect.~\ref{calcrossby}. 
In the first case the match between
the interior and the external layers is made at the optical depth $\tau$=2/3,
while in the non-gray treatment the boundary conditions are provided by
atmosphere models from \citet{allard00}, and a self-consistent integration is
performed down to an optical depth $\tau$=10.

\begin{figure}[tb]
\centering{
\includegraphics[width=9cm]{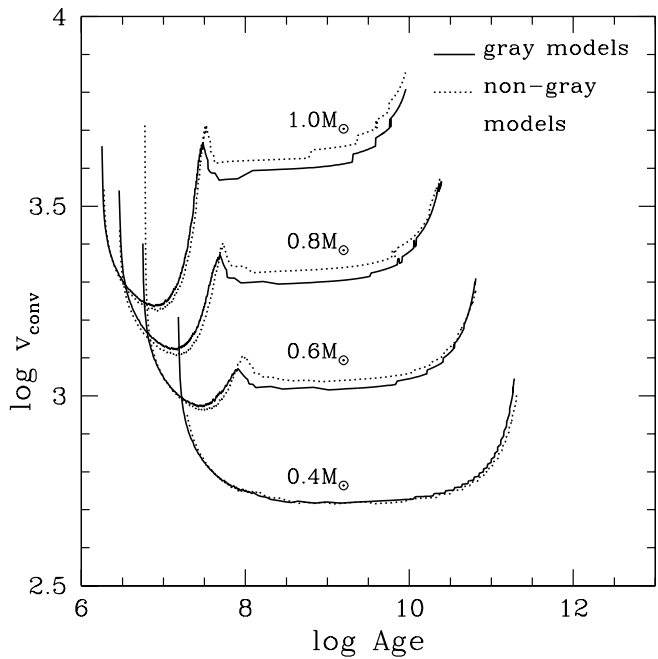}
\caption{Time evolution of the convective velocity $v$ for selected models
of 1.0, 0.8, 0.6 and 0.4 M$_\odot$ (from top to bottom) with non-gray and gray atmosphere
boundary conditions.}
\label{vconv}
}
\end{figure}

The convective velocities $v$ of non-gray models for the mass range considered
show an interesting behavior when compared to the corresponding ones of gray models, they are
lower during the pre-main sequence but are becoming higher as soon as they reach
the zero-age main sequence (Fig. \ref{vconv}). This can be easily understood if one
takes in account that, as shown by \citet{harris06,harris07} and now confirmed
by our own models, the use of non-gray atmospheres results in effectively larger opacities
and lower effective temperatures during the pre-main sequence evolution of low-mass stars.
Since a larger opacity favors higher convective velocities (see e.g. \citealt{maeder09})
but a lower temperature does the opposite, our models shows that this latter effect
prevails over the former during the pre-main sequence. From the zero-age main sequence on
the situation reverses, as the temperatures become nearly the same for both non-gray
and gray atmosphere models, but the opacity remains larger for the non-gray models,
and so results in higher convective velocities. 

\begin{figure}[tb]
\centering{
\includegraphics[width=9cm]{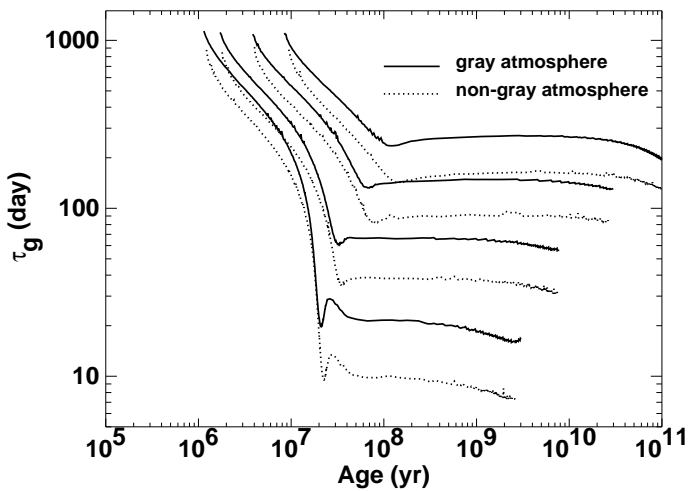}
}
\caption{Global convective turnover time as a function of age 
for 0.5, 0.7 1.0 an 1.2 M$_{\odot}$ models (from top to bottom) 
and for different atmospheric treatment.
}
\label{taugngatm}
\end{figure}

These general effects on the convective turnover time can be seen in Fig.~\ref{taugngatm}, where $\tau_{\rm g}$
is plotted against the stellar age for 0.5, 0.7, 1.0 and 1.2 M$_{\odot}$ models
computed with gray (solid lines) and non-gray (dotted lines) atmosphere boundary conditions.
Curves at the top of the figure correspond to the 0.5\,M$_{\odot}$, and the global convective
turnover time decreases as the stellar mass increases.

\section{Applications and comparisons with observations}\label{apply}

Star cluster data have particular importance in testing 
evolutionary models with rotation. The spin-down of young cluster stars places
a lower limit to the time scale for the angular momentum redistribution at
the base of the surface convection zone. Transport of the angular momentum leads to
rotationally induced mixing which reproduces the observed Li depletion in the 
Sun and solar analogs in open clusters \citep{pinsonn89}.
Observations of rotation rates of stars in clusters of different ages
help us to draw a scenario for the stellar angular momentum evolution.
The broad period distribution of zero-age main-sequence stars and the 
existence of slow rotators can only be explained if during early phases of
their evolution some stars lose a significant fraction of their angular momentum
which in addition must be different for each star (\citealt{lamm05}).  
A number of studies have been published in the literature showing how
observational data from open clusters can impose constraints on stellar models
as for example \citet{landin06} and \citet{rodriguez09} for the Orion nebula cloud
or \citet{meibom09} for M35.

In this context, and to give a simple application to our results, we used the \texttt{ATON} code to
calculate Rossby numbers for a representative sample of solar-type stars in the young 
(30 Myr) open cluster IC 2602, located at the southern hemisphere at a distance of
about 150 pc. 

As can be recalled from the introduction, the Rossby number is not a quantity directly obtained from 
observations, since it is the ratio of the local convective turnover time $\tau_c$ to the rotation
period $P_{\rm rot}$. While $\tau_c$ can be derived from evolutionary models or through a polynomial fit 
to B$-$V \citep{noyes84}, $P_{\rm rot}$ can be obtained observationally or computed
through rotating evolutionary models, though this last method is obviously model-dependent. 
In this section, we computed $Ro$ for IC 2602 stars in two 
ways:
(i) by using $\tau_c$ from our models and $P_{\rm rot}$ from observations and
(ii) by using both $\tau_c$ and $P_{\rm rot}$ calculated through our models.
We designate the $Ro$ calculated as described in case (i) as 
``semi-theoretical'' Rossby numbers and those computed as in case (ii) 
as ``purely theoretical'' Rossby numbers.

\begin{figure}[t]
\centering{
\includegraphics[width=9cm]{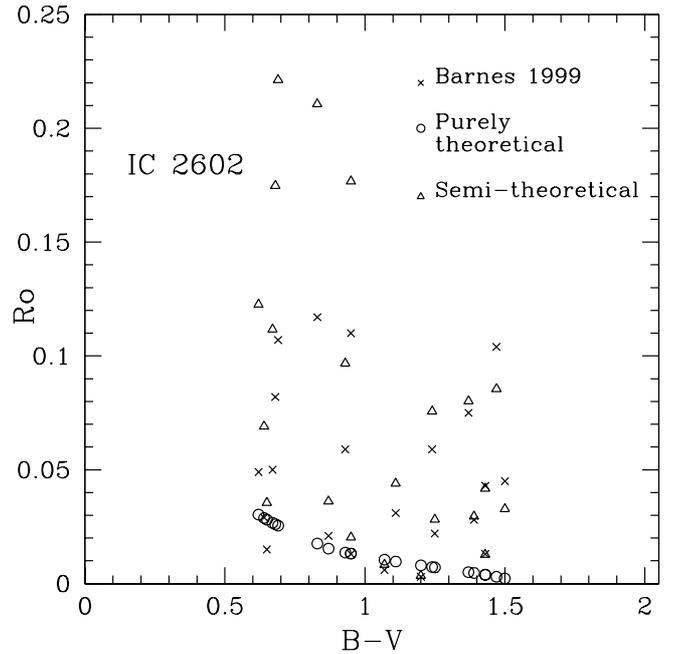}
}
\caption{Rossby number versus B$-$V color index for IC 2602 stars. Open
circles represent purely theoretical $Ro$, triangles stand for semi-theoretical
ones and crosses are $Ro$ presented by \citet{barnes99}. 
} 
\label{roic2602}
\end{figure}

The semi-theoretical values of $Ro$ were obtained 
by using the observed rotation periods of IC 2602 stars \citep{barnes99}, which are given
in the range from 0.2 days to $\sim$10 days as a function of the B$-$V color 
index, and $\tau_c$ calculated by our models at the age of 
IC 2602.
Since our models provide the local convective turnover time as a function of effective
temperature and gravity, we used the \citet{bessel98} relations,
which provide B$-$V for a grid of T$_{\rm eff}$ and $\log g$, to
obtain $\tau_c$ as a function of B$-$V.
In this way, by using $\tau_c(B-V)$ from the models and
$P_{\rm rot}(B-V)$ from observations, we calculated $Ro(B-V)$.
The purely theoretical Rossby numbers, in turn, were obtained from rotating models (SB+Diff) 
with gray atmosphere boundary conditions, MLT convection treatment ($\alpha$=1.5)
and internal
angular momentum redistribution with surface angular momentum loss, for the mass range of
0.2-1.4~M$_{\odot}$ at the age of IC 2602. Those models provide
us with $\tau_c$ and $P_{\rm rot}$ as a function of $T_{\rm eff}$ and gravity,
and, again by means of \citet{bessel98} relations, we obtained the purely
theoretical Rossby numbers corresponding to the observed B$-$V of IC 2602 
stars.  
In Fig.~\ref{roic2602} we present
our purely theoretical Rossby numbers (circles) and semi-theoretical ones 
(triangles) for the IC 2602 star sample. 
Crosses in Fig.~\ref{roic2602} represent the Rossby numbers presented by 
\citet{barnes99} and were plotted here for comparison purposes. These $Ro$
values were computed with the use of models from \citet{kim96} and are 
also designated as ``semi-theoretical'', since their way of computing 
$Ro$ is similar to ours.
 
For a better comprehension of 
differences in these $Ro$ calculations, we also display the corresponding values in 
Table\,\ref{tabic2602}, in which Col. 1 gives the identification of each star of the IC 2602 sample; 
Col. 2 the color index B$-$V; 
Col. 3 the purely theoretical Rossby number; Col. 4 the semi-theoretical
Rossby number; and Col. 5 the semi-theoretical Rossby number presented by 
\citet{barnes99}.  

When comparing our semi-theoretical Rossby numbers to those from
\citet{barnes99}, the differences are obviously due only to the corresponding
local convective turnover times, since the rotation periods are the same.
As discussed in Sect.~\ref{calcrossby}, these discrepancies
in $\tau_{\rm c}$ can be attributed to differences in the convection efficiency,
atmospheric boundary conditions, opacities and treatment of transport of
angular momentum.
Besides, although the distributions of our semi-theoretical
Rossby numbers and those from \citet{barnes99} are quite similar in Fig.~\ref{roic2602}
despite the differences of $\tau_{\rm c}$ presented in Table\,\ref{tabic2602}, one
must be cautious when interpreting those distributions.
\citet{barnes99} only mentioned that they used the $\tau_{\rm c}$ values from
the models of \citet{kim96} to calculate $Ro$ for the individual stars of their 
sample, but they do not disclose how they obtained their values of 
$\tau_{\rm c}$ as a function of B$-$V.

Our purely theoretical $Ro$ values, in turn, behave very 
differently from the semi-theoretical ones due to the different way 
in which rotation periods are obtained in these two cases.
In the first case the rotation periods start from an initial value of the
angular momentum given by Eq.\,(\ref{kaweq}) and evolve according to
local conservation of angular momentum in radiative zones and rigid body rotation in convective
ones; this initial value depends
on the stellar mass, i.e. also on B$-$V.
On the other hand in IC 2602, as in most young open clusters, there 
exist simultaneously ultrafast, intermediate and slow rotators 
independently of the stellar mass; in order to cover this broader period
distribution we need to consider some disk regulation mechanism
to describe the magnetic coupling of the central
star to its circumstellar disk. The role played by this disk regulation
in the rotational evolution of young clusters and its effects on Rossby numbers 
will be discussed in a forthcoming paper.

\begin{table}[ht]
\caption{Rossby numbers and B$-$V for IC 2602 stars.}
\label{tabic2602}
\centering
\begin{tabular}{lcccc}
\hline\hline
\multirow{2}{*}{Name$\,^{\rm a}$} & \multirow{2}{*}{B$-$V$\,^{\rm a}$} & $Ro$ & $Ro$ & \multirow{2}{*}{$Ro\,^{\rm a}$} \\
                                  &                                    & \scriptsize{(Pur. Theor.)} & \scriptsize{(Semi-Theor.)} & \\ \hline
B134 & 0.95 & 0.0132 & 0.1767 & 0.110 \\ [-1.7pt]
W79  & 0.83 & 0.0176 & 0.2107 & 0.117 \\ [-1.7pt]
R15  & 0.93 & 0.0136 & 0.0967 & 0.059 \\ [-1.7pt]
R24A & 1.43 & 0.0039 & 0.0127 & 0.013 \\ [-1.7pt]
R27  & 1.50 & 0.0023 & 0.0328 & 0.045 \\ [-1.7pt]
R29  & 1.11 & 0.0097 & 0.0440 & 0.031 \\ [-1.7pt]
R43  & 0.95 & 0.0132 & 0.0203 & 0.013 \\ [-1.7pt]
R52  & 1.07 & 0.0105 & 0.0084 & 0.006 \\ [-1.7pt]
R56  & 1.43 & 0.0039 & 0.0418 & 0.043 \\ [-1.7pt]
R58  & 0.65 & 0.0281 & 0.0355 & 0.015 \\ [-1.7pt]
R66  & 0.68 & 0.0261 & 0.1748 & 0.082 \\ [-1.7pt]
R70  & 0.69 & 0.0255 & 0.2212 & 0.107 \\ [-1.7pt]
R72  & 0.64 & 0.0288 & 0.0691 & 0.029 \\ [-1.7pt]
R77  & 1.47 & 0.0030 & 0.0855 & 0.104 \\ [-1.7pt]
R83  & 0.62 & 0.0303 & 0.1226 & 0.049 \\ [-1.7pt]
R88A & 1.20 & 0.0080 & 0.0035 & 0.003 \\ [-1.7pt]
R89  & 1.24 & 0.0073 & 0.0757 & 0.059 \\ [-1.7pt]
R92  & 0.67 & 0.0267 & 0.1117 & 0.050 \\ [-1.7pt]
R93  & 1.37 & 0.0050 & 0.0802 & 0.075 \\ [-1.7pt]
R94  & 1.39 & 0.0047 & 0.0295 & 0.028 \\ [-1.7pt]
R95A & 0.87 & 0.0154 & 0.0362 & 0.021 \\ [-1.7pt]
R96  & 1.25 & 0.0071 & 0.0282 & 0.022 \\ [-1.7pt] \hline
\multicolumn{5}{l}{\scriptsize{$^{\rm a}$ \citet{barnes99}}}
\end{tabular}
\end{table}

\section{Conclusions}\label{conclusion}

Our results show the same trends for the Rossby number $Ro$ and the local convective turnover time $\tau_{\rm c}$ found
in the work by \cite{kim96} as, for example, the decrease of $\tau_{\rm c}$ during the
pre-main sequence phase and its nearly constant value from that point
on. Despite this, our values of $Ro$ are in general
lower by one order of magnitude, mainly due to
differences in the initial rotation rates. 

As already expected, $\tau_{\rm g}$, the global convective time, 
decreases as convection efficiency increases,
and consequently $Ro$ increases; the opposite situation occurs
when stars with masses larger than 1\,M$_{\odot}$ enter the main sequence.
We also found that the values of $\tau_{\rm g}$ obtained for FST models
are lower than those obtained with MLT ($\alpha$=1.0, 1.5 and 2.2), since the former
compute the convective velocities in an intrinsically different way, which describes
the whole spectrum of convective eddies.

The effect of rotation on $\tau_{\rm g}$ is less important for masses lower than
1~M$_{\odot}$, but for masses larger than this threshold the influence of 
rotation on $\tau_{\rm g}$ is more evident. Solid body (SB) rotation produces 
stonger deviations on $\tau_{\rm g}$ relative to the non-rotating value than
the SB+Diff case. By increasing the initial angular momentum, $\tau_{\rm g}$
tends to increase as well.

Among the effects analyzed here, the ones yielded by atmosphere boundary conditions 
are those that have the most influence on the values of $\tau_{\rm g}$. 
Models which use non-gray boundary conditions produce values of $\tau_{\rm g}$
which are lower than those of the gray ones.

Our models were applied to calculate Rossby numbers as a function of the B$-$V
color index for a sample of stars from the IC 2602 open cluster. Semi-theoretical values of $Ro$ calculated with our
$\tau_c$ are, on average, higher than those presented by \citet{barnes99}.
Purely theoretical and semi-theoretical Rossby numbers have a different
behavior due to the different origin of the rotation period which composes
$Ro$ in each case. 

\begin{acknowledgements}
The authors thank Drs. Francesca D'Antona (INAF-OAR, Italy)
and Italo Mazzitelli (INAF-IASF, Italy)
for granting them full access to the \texttt{ATON} evolutionary code.
We are also grateful to an anonymous referee for his comments and suggestions.
Financial support from the Brazilian agencies CAPES, CNPq and 
FAPEMIG is gratefully acknowledged.
\end{acknowledgements}


\begin{thebibliography}{} 

\bibitem [Alexander \& Ferguson, 1994] {alexander}
         Alexander, D.R., Ferguson, J.W., 1994, ApJ, 437, 879

\bibitem [Allard et al., 2000] {allard00} Allard, F., Hauschildt, P.H., 
         Schweitzer, A., 2000, ApJ, 539, 366

\bibitem [Barnes et al., 1999] {barnes99} Barnes, S.A., Sofia, S.,
         Prosser, C.F., Stauffer, J.R., 1999, ApJ, 516, 263
 
\bibitem [Bessel et al., 1998] {bessel98}
         Bessell, M.S., Castelli, F., Plez, B., 1998, A\&A, 333, 231

\bibitem [B\"ohm-Vitense, 1958] {bohm58} B\"ohm-Vitense, E.\ 1958, 
         Z. Astroph., 46, 108
 
\bibitem [Canuto \& Mazzitelli, 1991] {canuto91}
         Canuto, V.M., Mazzitelli, I., 1991, ApJ, 370, 295

\bibitem [Canuto \& Mazzitelli, 1992] {canuto92}
         Canuto, V.M., Mazzitelli, I., 1992, ApJ, 389, 724

\bibitem [Canuto et al., 1996] {canuto96}
         Canuto, V.M., Goldman, I., Mazzitelli, I., 1996, ApJ, 473, 550

\bibitem [Chaboyer \& Zahn, 1992] {chaboyer92} Chaboyer, B., Zahn, J.-P.,
         1992, A\&A, 253, 173
\bibitem [Chaboyer et al., 1995] {chaboyer95}
         Chaboyer, B., Demarque, P., Pinsonneault, M.H., 1995, ApJ, 441, 865

\bibitem [Christensen-Dalsgaard, 2008] {christensen08}
         Christensen-Dalsgaard, J., 2008, Mem. Soc. Astron. Ital., 79, 628

\bibitem [Dupret et al., 2004] {dupret04}
         Dupret, M.-A., Thoul, A., Scuflaire, R., Daszy\'nska-Daszkiewicz, J.,
         Aerts, C., Bourge, P.-O., Waelkens, C., Noels, A., 2004, A\&A, 415, 251
\bibitem [Durney \& Spruit, 1979] {durney79}
         Durney, B.R., Spruit, H.C., 1979, ApJ, 234, 1067

\bibitem [Durney et al., 1993] {durney93} Durney, B.R., De Young, D.S., 
         Roxburgh, I.W., 1993, Sol. Phys., 145, 207

\bibitem [Endal \& Sofia, 1976]{endal76} Endal, A.S., Sofia, S., 1976, ApJ, 210, 184

\bibitem [Endal \& Sofia, 1978]{endal78} Endal, A.S., Sofia, S., 1978, ApJ, 220, 279

\bibitem [Feigelson et al., 2003] {feigelson03} Feigelson, E.D., Gaffney, J.A.,
         Garmire, G., Hillenbrand, L.A., Townsley, L., 2003, ApJ, 584, 911

\bibitem [Flaccomio et al., 2003] {flacco03c} Flaccomio, E., Micela, G., 
         Sciortino, S., 2003, A\&A, 402, 277

\bibitem [Gilliland, 1986] {gilliland} Gilliland, R.L., 1986, ApJ, 300, 339

\bibitem [Harris et al., 2006]{harris06} Harris, G.J., Lynas-Gray, A.E., 
         Tennyson. J., 2006, Stellar Evolution at low Metalicity: Mass Loss,
         Explosions, Cosmology, edited by H. Lamers, N. Langer, T. Nugis 
         and K. Annuk, ASP Conference Series, Vol. 353

\bibitem [Harris et al., 2007]{harris07} Harris, G.J., Lynas-Gray, A.E., 
         Miller, S., Tennyson, J., 2007, MNRAS, 374, 337

\bibitem [Hauschildt et al., 1999] {allard1} Hauschildt, P.H., Allard, F., 
         Baron, E., 1999, ApJ, 512, 377

\bibitem [Heiter et al., 2002] {heiter} Heiter, U., Kupka, F., 
         van't Veer-Menneret, C., Barban, C., Weiss, W.W., Goupil, M.-J., 
         Schmidt, W., Katz, D., Garrido, R., 2002, A\&A, 392, 619

\bibitem [Iglesias \& Rogers, 1991] {rogers91} Iglesias, C.A., Rogers, F.J., 
         1991, ApJ, 371, 408

\bibitem [Iglesias \& Rogers, 1993] {rogers1} Iglesias, C.A., Rogers, F.J., 
         1993, ApJ, 412, 752

\bibitem [Jung \& Kim, 2007] {jung2007} Jung, Y.C., Kim, Y.-C.,
         2007, J. Astron. Space Sci., 25, 30

\bibitem [Kawaler, 1987] {kawaler87} Kawaler, S.D., 1987, PASP, 99, 1322

\bibitem [Kim \& Demarque, 1996] {kim96} Kim, Y.-C., Demarque, P.S., 1996, 
         ApJ, 457, 340

\bibitem [Kippenhahn \& Thomas, 1970]{kippen70} Kippenhahn, R.,
           Thomas, H.-C.\ 1970, in {\it Stellar Rotation},
           ed. A. Slettebak (Dordrecht: Reidel)

\bibitem [Kurucz, 1991]{kurucz91} Kurucz, R.L., 1991, in {\it Stellar 
         Atmospheres: Beyond Classical Models}, ed. L. Crivellari, 
         I. Hubeny, \& D. G. hummer (Dordrecht: 
         Kluwer), 440
 
\bibitem [Kurucz, 1992]{kurucz92} Kurucz, R.L., 1992, in {\it IAU Symp. 149,
         The Stellar Population of Galaxies}, ed. B. Barbuy \& A. Renzini,
         Kluwer Academic Publishers, Dordrecht, The Netherlands, 225

\bibitem [Lamm  et al., 2005]{lamm05} Lamm, M.H., Mundt, R., Bailer-Jones, C.A.L.,
         Herbst, W., 2005, A\&A, 430, 1005

\bibitem [Landin et al., 2006] {landin06} Landin, N.R., Ventura, P., 
          D'Antona, F., Mendes, L.T.S., Vaz, L.P.R., 2006, A\&A, 456, 269

\bibitem [Li et al., 2006] {li06}
         Li, L. H., Ventura, P., Basu, S., Sofia, S., Demarque, P., 2006, ApJS 164, 215
\bibitem [Lydon \& Sofia, 1995] {lydon95} Lydon, T. J., Sofia, S., 1995, ApJS 101, 357

\bibitem [Ludwig et al., 1999] {ludwig99} Ludwig, H., Freytag, B., 
         Steffen, M., 1999, A\&A, 346, 111

\bibitem [Ludwig et al., 2002] {ludwig2002} Ludwig, H.-G., Allard, F., 
         Hauschildt, P.~H., 2002, A\&A, 395, 99

\bibitem [Maeder, 2009] {maeder09} Maeder, A., 2009, in 
         {\it Physics, Formation and Evolution of Rotating Stars},
         Springer (Berlin)

\bibitem [Meibom et al., 2009] {meibom09}
         Meibom, S., Mathieu, R. D., Stassun, K. G., 2009, ApJ 695, 679

\bibitem[Mendes, 1999] {mendesphd} Mendes, L.T.S.\ 1999, Ph.D. Thesis,
           Federal University of Minas Gerais

\bibitem [Mendes et al., 1999] {mendes99} Mendes, L.T.S., D'Antona, F., 
         Mazzitelli, I., 1999, A\&A, 341, 174

\bibitem [Mendes et al., 2003] {mendes03} Mendes, L.T.S., Vaz, L.P.R., 
         D'Antona, F., Mazzitelli, I., 2003, Open Issues in Local Star 
         Formation and Early Stellar Evolution, edited by J. L\'epine, 
         and J. Gregorio-Hetem, Astrophysics and Space Science Library 299, 
         Kluwer Academic Publishers, Dordrecht, The Netherlands

\bibitem [Mohanty \& Basri, 2003] {mohanty03} Mohanty, S., Basri, G., 
         2003, AJ, 583, 451

\bibitem [Montesinos et al., 2001] {montesinos01} Montesinos, B., 
         Thomas, J.H., Ventura, P., Mazzitelli, I., 2001, MNRAS, 326, 877

\bibitem [Noyes et al., 1984] {noyes84} Noyes, R.W., Hartmann, S.,W., 
         Baliunas, S., Duncan, D.K., Vaughan A., 1984, ApJ, 279, 763

\bibitem [Pinsonneault et al., 1989]{pinsonn89} Pinsonneault, M.H.,
         Kawaler, S.D., Sofia, S., Demarque, P., 1989, ApJ, 338, 424

\bibitem [Pizzolato et al., 2001] {pizzolato01} Pizzolato, N., Ventura, P., 
         D'Antona, F., Maggio, A., Micela, G., Sciortino, S., 2001, 
         A\&A, 373, 597

\bibitem [Rodriguez-Ledesma et al., 2009] {rodriguez09}
         Rodríguez-Ledesma, M. V., Mundt, R., Eisl\"{o}ffel, J., 2009, A\&A, 502, 883

\bibitem [Skumanich, 1972] {skuma72} Skumanich, A., 1972, ApJ, 171, 565

\bibitem [Tang et al., 2008] {tang08}
         Tang, Y.-K., Bi, S.-L., Gai, N., Xu, H.-Y., 2008, Chin. J. Astron.
         Astrophys. 8, 421

\bibitem [Ventura \& Zeppieri, 1998a] {ventura98a} Ventura, P., Zeppieri, A., 
         1998a, A\&A, 340, 77

\bibitem [Ventura et al., 1998b] {ventura98} Ventura, P., Zeppieri, A., 
         Mazzitelli, I., D'Antona, F., 1998b, A\&A, 334, 953

\bibitem [Weiss \& Tobias, 2000] {weiss00} Weiss, N.O., Tobias, S.M., 
         2000, SSRv., 94, 99

\bibitem [Zahn, 1992]{zahn92} Zahn, J.-P., 1992, A\&A, 265, 115
\end{thebibliography}
\end{document}